Chapter 4

From Lacaille to Lalande: French work on lunar distances, nautical ephemerides and lunar tables, 1742–85 [FINAL DRAFT]

Guy Boistel

In the 1630s, when Galileo Galilei sought a longitude reward from the Dutch States for his curious floating telescopic device for observing eclipses of Jupiter's satellites, lunar distances were under intense discussion in France between Cardinal Richelieu and the French savant Jean-Baptiste Morin.[1] Drawing on the work of sixteenth-century authors including Gemma Frisius, Johannes Werner and Peter Apian, Morin set down 13 propositions outlining astronomical and computational methods for finding longitude from the Moon, including lunar distances, lunar altitudes, meridian transits and hour angles.[2] Morin also described the 'clearing' of observations for refraction and parallax.

As Parès has explained, while the true angular distance would be the main objective of computations in the mid-eighteenth century, for Morin it was just one step in the process of defining the Moon's coordinates. From Morin's point of view, longitude difference had to be obtained by comparing the estimated coordinates of the Moon, deduced from lunar altitude and distance observations, with those computed from astronomical tables and tabulated in almanacs. In the mid-seventeenth century, however, stellar positions were not precisely known; the lunar motions had not been solved (predictions still being subject to errors of 30 to 50 arcminutes);

instruments for measuring angles were not sufficiently accurate; and astronomers did not have precise and reliable timekeepers. The principles were understood; the instruments had to be improved. Nonetheless, Morin's contributions were well known to eighteenth-century astronomers, including Abbé Nicolas-Louis de Lacaille and Canon Alexandre-Gui Pingré, who were aware of his propositions and used them as the basis of their own methods.

At the end of the seventeenth century, two increasingly dominant maritime states, France and England, established royal observatories, each with the remit of helping to solve the problem of determining longitude at sea. In England, as in France, the Astronomer Royal was instructed 'to apply himself with the most exact Care and Diligence to the rectifying the Tables of the Motions of the Heavens, and the places of the fixed Stars, so as to find out the so much desired Longitude of Places for perfecting the Art of Navigation'.[3] In France, the national map was significantly refined by the astronomers Jean Picard and Philippe de La Hire by determining terrestrial longitudes from observations of the eclipses of the two first satellites of Jupiter, as proposed by Galileo.[4]

Like Britain, France initiated rewards and academic prizes for solving the longitude problem and improving navigation at sea. Two years after the passage of the British Longitude Act, the French regent Philippe d'Orléans suggested establishing rewards, although the promise was not fulfilled. In 1722, the Académie royale des sciences elected to use a bequest from Count Rouillé de Meslay to fund a biannual prize with special reference to navigation. Although this was arguably to have only minor impact on nautical astronomical research in France, the proposed and actual prizes meant that the French government needed experts to examine projects claiming rewards.[5] As this chapter explores, the different ways in which appointed experts and astronomers interpreteted their role was to have a significant impact on the development of the theory and practice of navigation in France and elsewhere. Their work was, in particular, to influence that of the British Astronomer Royal, Nevil Maskelyne.

**Expertise as a constraint on research in France**

Efforts to determine the shape of the Earth in the early eighteenth century had an explicit connection with astronomers' and seafarers' desires to improve navigation. Behind the debate between Cartesians and Newtonians over the new physics lay the desire to know the shape of the Earth for navigational purposes. Indeed, one direct consequence of the expedition to Swedish Lapland (1736–37) was to put astronomers and members of the Académie more firmly in charge of improving navigation. Shortly after Pierre-Louis Moreau de Maupertuis returned from Lapland, the Ministre de la Marine, Jean-Frédéric Phélypeaux de Maurepas, made him the ministry's adviser for nautical sciences, as 'official responsible for the improvement of navigation and of the Navy in all its forms'.[6] As part of his new role, Maupertuis was to publish textbooks for seafarers. These works, the astronomical content of which has to be read with this demand in mind, included the *Élements de géographie* (1740), *Discours sur la parallaxe de la lune* (1741), the very strange *Astronomie nautique* (1743) and the *Traité de la loxodromie tracée sur la véritable surface de la Mer* (1748).[7] The phrase 'very strange' is used to qualify the *Astronomie nautique* because it proved a complete failure. Written for shipwrecked sailors, it was organized around analytical methods primarily for measuring latitude using pen and paper saved from the wreck. As can be imagined, it never entered naval schools, and, as discussed below, Maupertuis's views can be seen to have hindered Lacaille's aspirations to disseminate the lunar-distance method.

Pierre Bouguer succeeded Maupertuis in 1745. Bouguer had begun his career as a professor of hydrography and mathematics before gaining fame as a mathematician and member of the 'Peru expedition', the second geodetic investigation of the shape of the Earth, which was sponsored by the Académie (1733–44). Bouguer was then the best expert the Ministre de la Marine could find for institgating improvements to the navy. As early as 1726, however, he had signalled his rejection of mechanical timekeeping for longitude determination during his travels to the Equator, and maintained this stance in his reports as expert adviser between 1749 and 1758.[8] Such an authoritative rejection of timekeepers was to hold sway at the Académie until news of John Harrison's sea watch (H4) reached France in the 1760s.

Following Bouguer's death in 1758, his position was split between the mathematician Alexis-Claude Clairaut and the astronomer Pierre-Charles Lemonnier, both experts on lunar tables and their nautical uses yet with differing views and methods.[9] Jérôme Lalande replaced Clairaut after his death in May 1765.[10] Given responsibility for improving the navy, these savants worked under the control of the Ministre de la Marine, without interference from their peers or the Académie. Improving nautical astronomy was considered a task not for naval officers but for the scientific elite: royal astronomers and members of the Académie. Many of the books by these savants, mathematicians and astronomers – Maupertuis, Bouguer, Clairaut, Le Monnier and Lalande – should, therefore, be read in the context of their role as 'official responsible for the improvement of navigation', giving a more consistent and deeper view of their scientific activity.

Examining their works in this light, it is possible to draw some conclusions about the nature of their expertise and influence. With some hindsight, for example, three of the five might be said to have had a negative influence on the development of scientific navigation in France.

- Maupertuis failed to answer the Ministre de la Marine's demands and, for a while, halted Lacaille's attempts to develop nautical astronomy and lunar methods;[11]

- Bouguer discouraged timekeeping research during the 1750s, despite knowing skilled clockmakers (the Lepaute and Leroy brothers) with relevant expertise;[12]
- Lemonnier disseminated old astronomical and nautical methods; he was not able (or did not wish) to follow new developments in Newtonian celestial mechanics, something that overshadowed his quarrels with Lacaille, Clairaut and Lalande (his ex-pupil).

There were deep disagreements and controversies within and between the main scientific academic organizations too: within the Académie (between factions, as well as individually between d'Alembert and Clairaut, Lemonnier and Lacaille, Lalande and Lemonnier, Lalande and Cassini III, and others); within the Académie de Marine in Brest (Lemonnier was excluded in 1771, for example); and between successive Secrétaires d'état de la Marine because of an overly mercurial and hesitant French naval policy (in particular under Maurepas, Rouillé, Choiseul, Praslin, de Boynes and de Sartines).[13]

**The first observations of lunar distance at sea: Abbé Lacaille and Jean-Baptiste d'Après de Mannevillette, 1749–51**

After the passage of the Longitude Act in 1714 and the partial development of lunar theory by Isaac Newton, it seemed clear to astronomers that the Moon was the only natural clock that could be used regularly at sea. In 'A proposal of a method for finding the Longitude at sea within a degree' in the *Philosophical Transactions* for 1731, Edmond Halley offered a method based on observations of occultations of a star by the Moon for correcting lunar tables and calculating the ecliptic longitude of the Moon to within two arcminutes. In principle, the method was sufficiently precise for lunar methods at sea, and Halley's paper showed how the observation of a single angular distance between the Moon and a fixed star could help the seafarer determine longitude. Years later, Lacaille and the Jesuit professor of hydrography and astronomer in Marseille, Father Esprit Pezenas, would write that Halley's method was merely a variant of Morin's lunar altitude method, in which the lunar distance was only an intermediate step towards calculating longitude, not the endpoint.[14] Halley's paper, and the publication in 1742 of his *Astronomical Tables* with the corrections of lunar tables from observations of a Saros, would also play an important part in Lemonnier's and Pingré's longitude developments, discussed below.[15]

In 1742, Nicolas-Louis de Lacaille, 'adjoint astronome' of the Académie, took charge of the computation and publication of the *Ephémérides des mouvements célestes*. These were computed for a period of ten years and gave the astronomical data needed to compute calendars. Lacaille was aware of the recent requests for a renewal of the methods of nautical astronomy and of Halley's 1731 paper. In 1742, while working on the fourth volume of the *Ephémérides* for 1745–55, Lacaille thought of adding considerations of new ways of finding longitude at sea from lunar distances. Hearing that Maupertuis, his superior at the Académie and the official 'responsible for the improvement of the navigation', was shortly to publish the *Astronomie nautique* (1743), however, he shelved his plan, assuming that Maupertuis' book would deal with longitude at sea. As it turned out, it did not.

A new opportunity arose, however, when Lacaille met Jean-Baptiste d'Après de Mannevillette, an officer of the Compagnie des Indes who was based in the port of Lorient in Brittany and had good relationships with instrument makers in Paris. In June 1749, having made improvements to the octant, Mannevillette was the first naval officer to apply the lunar-distance method at sea, near Cape Verde, using an octant of Caleb Smith's design.[16] Well trained in mathematics, Mannevillette later said that he was able to determine longitude with an accuracy of between five and 15 'lieues marines' or marine leagues (25 to 45 km); in other words, to an accuracy greater than that required by the 1714 Longitude Act.[17] His results were published much later, in 1775, in the *Neptune François* collection of sea charts.[18]

Mannevillette's voyage to the Cape of Good Hope (1750–54) with Nicolas-Louis de Lacaille helped rekindle practical interest in the lunar-distance method. Both used the method to determine the longitude of Santiago in the Cape Verde Islands with considerable accuracy in November 1750. Several determinations of longitude differences were also made by lunar distance (from Antares) in Rio de Janeiro in January 1751. Given his skill in determining stellar positions, improving tables of atmospheric refraction and correcting tables for solar motions, not to mention his familiarity with Clairaut's work on lunar theory, it is no surprise that Lacaille was able to deploy and correct the tables required for carrying out the lunar-distance method. After completing his work on lunar parallax, geodesy and stellar cataloguing in Cape Town, Lacaille developed his ideas on the voyage back to France. In 1754, he sent a memorial on his new method to the Académie.[19] Noting that most seafarers lacked the scientific training to carry out the lunar method, he argued that it should be adapted and put 'within the reach of ordinary sailors':

> During this sea voyage, I occupied myself in making trials of the method of observing longitude at sea by means of the distances of the Moon from some zodiacal fixed star. Following my departure from France, I made numerous investigations to facilitate the practice of the method proposed by Mister Halley. I recognized that it was useless to look for another way of using the Moon for the longitude; that it was solely a question of making the calculation easy for ordinary sailors. [20]

Lacaille also proposed computing the predicted lunar distance from the Sun and other key stars every three hours in a 'nautical almanac', the model Maskelyne would apply ten years later.

Nevertheless, there was no consensus within the Académie. Lacaille's main rival, the astronomer Pierre-Charles Lemonnier, was attempting to publish the first nautical almanac entirely devoted to the lunar altitude method (as well as that of hour angles).[21] The computations were performed by Alexandre-Guy Pingré. The resulting almanac, the *État du ciel*, was published four times between 1754 and 1757, but does not seem to have been used by seafarers owing to its complex computations. Nonetheless, with this publication in progress and with Lacaille at sea near Isle de France, Lemonnier was able to stop Lacaille's project.

**Lacaille's graphical method**

After his return to France in 1755, Lacaille was able to defend his proposal. He read a memorial to the Académie in 1759, which set out his plan for a pre-calculated table of lunar distances and added a graphical method to avoid the long and difficult calculations normally required by the lunar-distance method. His ideas were later promoted by Jérôme Lalande, who was elected in 1759 to take charge of the computation and publication of the *Connaissance des temps* (hereafter *CDT*). Lalande had some original views on the *CDT* and its contents, notably adding new scientific matter that can often now only be found there. The volume for 1761, for instance, included Lacaille's procedures and methods for lunar distances. In 1755, he also published the *Ephémérides des mouvements célestes*, astronomical tables for ten years (1755–65), in which he gave further consideration to lunar distances and his longitude method. In 1760 Lacaille also published a revised edition of Bouguer's *Nouveau traité de navigation*, which expanded on his graphical method for lunar distance, a method Bouguer had previously overlooked.

Lacaille's graphical method derived from an idea proposed in 1692 by Father Paul Hoste, S.J., professor of hydrography for the navy in Brest, and explained more carefully by Bouguer in the first edition of the *Nouveau traité de navigation* (1753). The method was also taught in Saint-Malo by a navigation teacher called Griffon, for whose 1748 memorial to the Académie Lacaille was the academic referee.[22] Griffon proposed a developed version of Hoste's method. The basis was to draw a circle representing the celestial sphere with the observer at the centre, then plot the pole and the equator at right angles to each other. To determine local time, it was easy to plot the Sun's path, the only data required being solar declination, which mariners could look up in almanacs. Following Bouguer's elaboration, with worked examples, in the *Nouveau traité de navigation*, Lacaille extended it to the determination of the apparent angular distance between the Moon and Sun. In doing so, he transformed the computations of spherical trignonometry into graphical operations of simple geometry, using only ruler, compass and the four basic operations, something easily within the grasp of the common seafarer.[23] Lacaille did not, however, specify the elements needed to clear or correct the angles for lunar parallax.

For his voyage to St Helena in 1761, Nevil Maskelyne took the *CDT* for 1761 and Lacaille's ephemeris. In a letter published in the *Philosophical Transactions* the following year, he discussed Lacaille's method, in particular the graphical method, agreeing with Lacaille on one point: the practical dispositions needed for the lunar distance method. Like Lacaille, he felt it best to have three observers measuring the two altitudes and the angular distance simultaneously, thus avoiding the need to calculate by interpolation the small but significant horary motion of the Moon, which would be a source of error.[24] As Sadler has explained, Maskelyne was unable to use the graphical method because of an error Lacaille made in the example calculations.[25] In fact, the error came from Lalande in the *CDT*s for 1761 and 1762, which had mistakenly swapped the figures for Regulus and Aldebaran in the examples for 8 July 1761.[26]

Like the Chevalier Jean-Charles de Borda after him, Maskelyne thought that Lacaille was too pessimistic about the accuracy with which one could measure the angular distance between the Moon and a star.[27] Nonetheless, Lacaille was the first astronomer to study the propagation of errors, drawing on Roger Cotes's *Harmonia Mensurarum* (edited by Robert Smith, 1722). Lacaille believed an accuracy of about four minutes of arc was possible; Maskelyne and Borda gave one minute (of arc) for the angular distance, and preferred to develop the lunar-distance method without simplified techniques.

In fact, Borda's opposition to Lacaille's methods went deeper. As already noted, Lacaille had argued that lunar distances had to be be adapted to seafarers' use, a view echoed in Alexis Rochon's and Lalande's beliefs that the astronomer's task was to simplify and popularize:

> We have simplified through tables all the other parts of the longitude calculations […] This part, however, greatly lengthens the longitude method and prevents many seafarers from engaging in these studies: if they [seafarers] continue to neglect these observations at the risk of their fortunes and lives, it is the astronomers' duty to lessen the difficulties and to remind them of the vital matters at stake.[28]

In the same year as Rochon's memorial, however, Borda condemned the use of graphical methods, which had 'the drawback of having men, only too inclined to it, becoming used to a process in some way automatic'[29]. Borda concluded elsewhere that the best way for navigators to avoid the difficulties and inconveniences of calculation was to be properly taught how to calculate. Borda's expectations were high, his opinion of seafarers low: 'It is about time that seafarers ceased looking at the mathematical and physical sciences as having no practical use in navigation and its progress. Without the help of science, the Navy would still be in its infancy'.[30] In the eyes of many mathematicians, examiners and savant-officers of the French navy, it was up to seafarers to rise to the demands required by the new navigational methods, rather than for mathematicians to simplify solutions and contrivances to circumvent direct calculation.[31]

**French and Portugese attempts to publish nautical almanacs**

The development of lunar methods for determining longitude at sea was directly connected to the establishment of nautical almanacs, not only in France, but also in Portugal. In France, the process began in the early 1750s with a dispute over the contents of the *CDT*. The Abbé André-François Brancas de Villeneuve proposed modifying the *CDT* to transform it into a nautical almanac, and published *Éphemérides cosmographiques* between 1750 and 1755. In April 1755 he sent a memorial to the Ministre de la Marine and the astronomer Joseph-Nicolas Delisle – one of Lalande's former teachers – in which he castigated the astronomers and exorted them to concentrate on their real work: producing tables for longitude determinations.[32] Brancas added that such a nautical almanac should be published two or three years in advance. While his other proposals were sometimes idiosyncratic, Brancas set out all the principles, known from Delisle and Lalande, for creating a nautical almanac.[33] At the same time, Lacaille and Lemonnier were working on proposals for new nautical ephemerides to help seafarers. Nor were these the only attempts to publish nautical almanacs in France. In several ports, small nautical almanacs existed, called (with local variations) *Étrennes maritimes et curieuses, Étrennes nautiques, Étrennes nantaises*, and similar. These gave the times of rising and setting and the declinations of the Sun and Moon, information needed to determine local time from the altitude of either body; in other words, the basic astronomical elements used in Hoste's and Bouguer's graphical methods.

Likewise, in 1758 the Portuguese Jesuit Eusebio da Veiga published the *Planetario Lusitano,* a type of nautical almanac, a year before the dispersal of the Jesuit order.[34] This latter event was to disrupt maritime scientific education for several years in France, and more in Portugal, since maritime education in both countries was mainly conducted by Jesuits. During the 1750s, therefore, there was significant activity around the problem of improving existing ephemerides in France and Portugal and of encouraging astronomers in the belief that they should produce the necessary tools for finding longitude at sea.

**Lalande, Jeaurat, Maskelyne and lunar distances in the *Connaissance des temps***

Elected as the new director of the *Connaissance des temps* in 1759, Lalande worked over the next deacde or so to develop lunar-distance methods, despite considerable resistance from the Académie. In 1772, for example, the Académie de marine in Brest proposed translating the British *Nautical Almanac* and publishing an edition of Dunthorne's and Witchell's formulae for clearing lunar distances. Initially, however, the Ministre de la Marine refused to grant the privilege for printing and forbade any translation, considering it an *allégence* to France's maritime rival; there was also a question of rivalry between the two royal academies. Only after several exchanges with the Ministre did Lalande, who was a member of both academies (Paris and Brest), gain authorization to add lunar-distance tables to the *CDT*. As Jim Bennett has shown, there was only a weak notion of state secrecy regarding longitude discoveries.[35] There were no secrets regarding astronomical and nautical computations, with such information circulating easily and quickly between Maskelyne and French astronomers at the end of the eighteenth century.[36]

Lalande and his pupils Edme-Sébastien Jeaurat and Pierre Méchain used the work of Maskelyne's computers to complete French ephemerides between 1772 and 1785, as well as lunar-distance tables reduced to the Greenwich meridian, before beginning to compute the same tables, reduced to the Paris meridian after 1790, with the help of the first ever full-time lunar-distance computer, Louis-Robert Cornelier-Lémery.[37] The need for purely French nautical ephemerides became even more pressing when Lalande discovered in 1803 that Maskelyne had made errors of five to six seconds in the positions of the stars needed for lunar-distance and other tables:

> We are occupied these days in recalculating from observations Maskelyne's 34 stars which we've used with complete confidence, and I find it is necessary to add 5 or 6 seconds to the right ascensions. So we'll have to correct all our catalogues, all our tables and all our longitudes of the observed planets! This old pen pusher, lazy drunkard, miser, has usurped our trust. He's very rich, he should have got himself a computer and checked, more than once, this important result.[38]

**Lunar tables for longitude: Mayer versus Clairaut, or empiricism versus theory?**

To understand how theoretical knowledge circulated within the network of European astronomers and *géomètres* (mathematicians), it is necessary to examine the ways in which astronomical tables of the Moon's motions were developed between 1750 and 1770 as the basis for computations of the lunar-distance elements of nautical almanacs and ephemerides. In this context, it is important to remember that in 1765 a reward was given posthumously to Tobias Mayer for his lunar tables. To achieve an accuracy of half a degree of longitude, as specified by the Longitude Act, the tables had to be able to give the ecliptic longitude of the Moon to within one arcminute.

Less well known than the award to Mayer's widow is a letter (in English) of 11 April 1765 from Alexis Clairaut to John Bevis, claiming an equal part of the reward.[39] Clairaut was not generally known as a mathematician involved in the development of nautical astronomy, nor for his commitment to the improvement of the navigation, yet his letter claimed that his lunar tables were superior to Mayer's. There was some history to this work. Despite repeated efforts, Newton had failed to formulate a complete theory of the Moon's motions, leaving to later mathematicians and astronomers the task of solving by approximation the three-body problem; that is, the Keplerian problem of the motions of two celestial bodies, but also taking into account the perturbations caused by a third body. In fact, this problem has no analytical and exact solution and can be solved only by successive approximations, the theory of perturbations. The Moon's ecliptic longitude is obtained by the addition of terms which appear as smaller and smaller corrections to the elliptical Keplerian orbit. From 1743 Clairaut, Euler and d'Alembert developed such a theory in an atmosphere of intellectual competition and rivalry over the motions of both the Moon and comets.[40]

But how did these *geomètres* construct the tables from which astronomers might compute navigational terms and, most importantly, the lunar distances published in nautical ephemerides? And how did astronomers correct or adjust the theoretical computations of the Moon's position against observed lunar positions? Eighteenth-century *geomètres*, mathematicians and/or astronomers had very different, and often unambiguous, opinions on these questions. On one hand, some argued, tables had to be obtained only from theory, a pure theory. On the other, it seemed necessary to make corrections of the Moon's position by reference to practical observations: the theory was modified, for example, with a term obtained from the computation of the mean of O – C (observed minus computed). In other words, the result could be seen as empirical. Clairaut developed his tables with 22 terms obtained theoretically; Mayer developed his with 26 terms corrected from observations.

The divergence between theoretical and empiricial developments in celestial mechanics seems to have begun when Bouguer wrote of the empiricism of Mayer's lunar tables in 1754. For Bouguer, the correction of theoretical terms by means of observations could only be a short-term solution and was intellectually unsatisfactory. Clairaut followed the same line, which he repeated in his letter to Bevis, writing that:

> as I have done it by the meer theory, it is to hope that their agreement with the observations will hold more constantly than that which is grounded upon an empirick method, which may be good for a time not very distant from the observations made use of in the confection of the tables, and disagree afterwards.[41]

For Clairaut and d'Alembert, the correction of lunar tables by means of observations was not the right way to develop lunar theory:

> We can even observe that in the equations M. Mayer uses for his tables, the values of the coefficients are not exactly the same as those he extracted from theory; from which it would appear that the tables of M. Mayer were partly drawn up from observations, by a sort of trial-and-error method, combined with the principal results derived from theory.[42]

Leonhard Euler wisely noted in 1765 that the three-body problem was highly complex and that mathematicians needed time to complete the project; it was premature to argue about it. Even when Euler's son, Johan Albrecht, wrote in 1766 that the lunar tables produced by Clairaut and Mayer were sufficiently accurate for calculating lunar distances, he added that the problem was still not really solved; the necessary approximations and theory remained incomplete.

A number of French astronomers helped Clairaut with his calculations; apparently both he and d'Alembert disliked number crunching. Delisle performed the calculations for the lunar tables published in 1754; in 1763–64, Bailly, Jeaurat and Pingré did the same for the second edition of Clairaut's theory of the Moon's motions (Saint-Petersburg, 1765) and for the 1764 annular eclipse of the Sun. Clairaut's and Mayer's lunar tables were also tested in 1764 based on their predictions of an annular eclipse of the Moon due to occur on 1 April; Mayer's tables suggested that the eclipse would not be seen in Paris, Clairaut's that it would. Clairaut won this test, since the eclipse was indeed observed in Paris. The astronomers Cassini III, Bailly and Pingré subsequently recommended that the Académie compute the lunar elements published by Lalande in the *CDT* on the basis of Clairaut's tables, considered as 'pure' theoretical tables. The rejection of Mayer's tables by some French astronomers, Lacaille and Lalande aside, can also be understood in the light of his failure to answer Lacaille's (and d'Alembert's) challenge of explaining the fundamentals of his theory of lunar motion.[43]

Most French astronomers defended and used Clairaut's lunar tables until the late 1780s. Edme-Sébastien Jeaurat began comparisons between Clairaut's and Mayer's tables in June 1759, completing major studies for 1759, 1764, 1776–79, 1780 and 1781–82, the last two also performed by Cornelier-Lémery. In the first three studies, Clairaut's lunar tables came out well: the discrepancies of the errors (O – C) were similar to those of Mayer's tables. Both sets of tables proved to be accurate, with a mean error about one minute of arc for the ecliptic longitude of the Moon. At the beginning of the 1780s, however, Jeaurat and Lémery pointed out that the discrepancies in Clairaut's lunar tables were increasing because, unlike his rivals, Clairaut had not included the secular acceleration of the Moon. Subsequently, Lémery mainly used Euler's tables, computed from his second theory of the lunar motions, until the beginning of the nineteenth century, when they were superseded by the tables of Bouvard, Bürckhardt, Damoiseau and Plana, all based on Laplace's celestial mechanics.[44]

**Conclusion**

The development of theories and practices for finding longitude at sea by lunar methods followed different courses in Britain and France in the mid-eighteenth century. British astronomers mainly focused on Halley's methods for lunar distances until the publication of the *Nautical Almanac* in 1767. French astronomers, by contrast, were aware of Jean-Baptiste Morin's methods and explored them in the 1740s and 1750s. From the 1750s, there were also attempts in France to develop and adapt ephemerides and nautical almanacs for the needs of seafarers. Lacaille played an important and significant role in this. He developed the lunar-distance method and gave a model for a nautical almanac containing pre-computed angular distances between the Moon and a bright star every three hours. His works were a source of inspiration for Maskelyne in Britain.[45] Lalande promoted Lacaille's graphical method in the *Connaissance des temps*. Not fully explained, either by Lacaille or by Lalande, this method was tried and discussed by Maskelyne during his sea trials, but was abandoned. Nonetheless, the need expressed by Lacaille to develop simplified methods for 'ordinary sailors' began to be met by Jérôme Lalande and Abbott Alexis Rochon at the end of the eighteenth century.

Nevil Maskelyne had certainly read Lacaille and Lalande by the start of the 1760s, but what did he know of Lemonnier and Pingré's attempts to produce the *État du Ciel* ephemeris? What did he know of French debates on longitude? And what influence did they have on Maskelyne's own thinking about these problems and their solutions? Maskelyne's journal of his voyage to St

Helena between in 1761 offers some evidence. On page after page, his work shows knowledge of celestial mechanics, mainly from Clairaut's early works and Mayer's 1755 tables, and draws from them every element concerned with the problem of determining longitude at sea: angular distance, clearing for the effects of parallax and refraction.[46] Maskelyne had clearly read Lacaille and Lalande in the *Connaissance des temps*, as well as Bouguer's *Nouveau traité de navigation* as revised and expanded by Lacaille in 1760. The *Nautical Almanac* appears, therefore, to be an adaptation of the French ephemeris and a realization of Lacaille's earlier proposals for a nautical almanac. Nonetheless, the full extent of what Maskelyne owed to Lacaille has yet to be fully explored.

---

[1] Jean Parès, 'Jean-Baptiste Morin (1583–1654) et la querelle des longitudes de 1634 à 1647' (Thèse de Doctorat de 3ième cycle, Paris, E.H.E.S.S., Université Paris-I, 1976).

[2] Jean-Baptiste Morin, *La science des longitudes de Jean-Baptiste Morin* (Paris: aux dépens de l'Autheur, 1647), pp. 21–40, for Morin's 13 propositions.

[3] Royal warrant, quoted in Derek Howse, *Greenwich Time and the Longitude* (London: Philip Wilson/National Maritime Museum, 1997), p. 42.

[4] *Jean Picard et les débuts de l'astronomie de précision au XVII$^e$ siècle*, ed. by Guy Picolet (Paris, CNRS Éditions, 1987). See also, J. D. North, 'The Satellites of Jupiter, from Galileo to Bradley', in *The Universal Frame. Historical Essays in Astronomy, Natural Philosophy and Scientific Method* (London: Hambledon Press, 1989), pp. 185–214.

[5] Guy Boistel, 'L'astronomie nautique au XVIII$^e$ siècle en France: tables de la Lune et longitudes en mer' (Thèse de doctorat en épistémologie, histoire des sciences et des techniques, Centre François Viète, Université de Nantes, 2001), part I, pp. 30–75; Guy Boistel, 'Pierre-Louis Moreau de Maupertuis: un inattendu préposé au perfectionnement de la navigation (1739–

*Cahiers d'histoire et de philosophie des sciences*, 54 (2005)), 27–45; Boistel, 'L'astronomie nautique', part III.

[15] The name was given by Halley from the Chaldean cycle of 223 lunar periods. After each Saros cycle, the irregularities of Moon's motions are supposed to be the same. This cycle was used to compute eclipses of the Sun by the Moon. For the origins of the word *saros* and the uses of the cycle for the correction of lunar tables by Halley, Legentil de la Galaisière, Lemonnier, Pingré and d'Alembert in particular, see Boistel, 'L'astronomie nautique', part III, pp. 299–302 and part IV, pp. 508–70.

[16] A Caleb Smith quadrant appears in 'Portrait of a Merchant Captain', by Robert Willoughby, oil on canvas, 1805, National Maritime Museum, Greenwich, BHC3130.

[17] Jean-Baptiste d'Après de Mannevillette, 'Principes du calcul astronomique, provenant du cabinet de d'Après de Mannevillette' (*c*.1754), Fonds du Dépôt des Cartes et Plans, Service Historique de la Défense, Paris, Vincennes, SH 53.

[18] See Boistel, 'L'astronomie nautique', part III, pp. 281–500 for a study of Lacaille's and Mannevillette's works, longitude determinations at sea and methods; see also *Autour de d'Après de Mannevillette: savant navigateur havrais du siècle des Lumières*, ed. by Eric Saunier (Le Havre: Centre Havrais de Recherche Historique, 2008); Manonmani Filliozat, 'D'Après de Mannevillette, capitaine et hydrographe de la Compagnie des Indes (1707–1780)' (Thèse de l'École des Chartes, Paris, 1993).

[19] Nicolas-Louis de Lacaille, 'Projet pour rendre la méthode des longitudes sur mer pratiquable au commun des navigateurs', 1754, A.N., Marine, 2 JJ 69 (J.-N. Delisle's papers), item 16; Nicolas-Louis de Lacaille, 'Instruction détaillée pour l'observation et le calcul des longitudes sur mer par la distance de la Lune aux étoiles ou au Soleil', 1754, A.N., Marine, 3 JJ 13, items 3 and 9.

[20] Nicolas-Louis de Lacaille, 'Relation abrégée du Voyage fait par Ordre du Roi au cap de Bonne-Espérance, par M. l'abbé de la Caille ', *Histoire et Mémoires de l'Académie Royale des Sciences… pour l'année 1751* (Paris: Imprimerie royale, 1755), *Mémoires*, read on 13 November 1754, pp. 519–36 ; see also *Procès-verbaux des séances de l'Académie Royale des Sciences*, 13 November 1754, Tome 73, fols 475–93 (fol. 491): 'Je m'étais beaucoup exercé à ces sortes d'observations, et j'avais reconnu qu'il étoit inutile d'avoir recours à une autre façon d'employer la Lune pour les longitudes; qu'il ne s'agissait uniquement que d'en rendre le calcul pratiquable au commun des marins'.

[21] Boistel, 'L'astronomie nautique', part II, pp. 158–60; part III, pp. 386–412.

[22] 'Mémoire présenté par le Sieur Griffon', 23 August 1748, Archives de l'Académie des sciences, Paris, Quai Conti, Pochette de séance.

[23] Boistel, 'L'astronomie nautique', part III, pp. 479–500.

[24] Nevil Maskelyne, 'A Letter from the Rev. Nevil Maskelyne M.A., F.R.S., to Rev. Thomas Birch, D.D., Secretary to the Royal Society, Containing the Results of Observations of the Distance of the Moon from the Sun […] in Order to Determine the Longitude of the Ship from Time to Time', *Philosophical Transactions,* LII (1761), 558–77; Donald H. Sadler, 'Lunar Distances and the *Nautical Almanac*', *Vistas in Astronomy*, 20 (1976), 113–21; see also Boistel, 'L'astronomie nautique', part III and part IV for the computations of the horary motion of the Moon in the tables of Clairaut, Mayer and d'Alembert.

[25] Sadler, 'Lunar Distances'.

[26] Jérôme Lalande, *Connaissance des temps pour l'année 1761* (Paris: Imprimerie royale, 1759), pp. 192–93.

[27] Guy Boistel, 'De quelle précision a-t-on réellement besoin en mer? Quelques aspects de la diffusion des méthodes de détermination astronomique et chronométrique des longitudes en mer en France, de Lacaille à Mouchez (1750–1880)', *Histoire & Mesure*, XXI (2006), 121–56.

[28] *Histoire générale des Mathématiques par M. Montucla*, ed. by Jérôme Lalande, 4 vols (Paris: H. Agasse, 1799–1802), IV, 581: 'On a simplifié par des tables toutes les autres parties des calculs de la longitude […] Cependant cette partie allonge beaucoup la méthode des longitudes et empêche beaucoup de navigateurs de s'occuper de ces recherches ; s'ils négligent encore ces observations au risque de leurs fortunes et de leurs vies, c'est le devoir des astronomes de leur aplanir les difficultés et de les rappeler à de pressants intérêts.' See also, Alexis Rochon, *Mémoire sur l'astronomie nautique* (Paris: Imprimerie de Prault, 1798).

[29] "Elles [les méthodes graphiques] ont l'inconvénient d'habituer à un travail en quelque sorte automatique, des hommes qui n'y sont déjà que trop disposés" . REF : Borda & Lévêque, 1798 (an VII), (ref note 30), p. 472.

[30] Le Chevalier de Borda and Pierre Lévêque, 'Rapport sur le mémoire et la carte trigonométrique présentés par le Citoyen Maingon, Lieutenant de Vaisseau', *Procès-Verbaux de l'Académie des sciences de l'Institut de France*, I, séance du 11 Vendémiaire an VII (2 October 1798) (Paris: Gauthier-Villars, 1910), pp. 465–73 (p. 473): 'Il est temps que les marins cessent de regarder les sciences mathématiques et physiques comme inutiles à la pratique de la navigation et à ses progrès. Sans le secours des sciences, la Marine seroit encore dans l'enfance'.

[31] For more detailed studies, see Boistel, 'De quelle précision'; Guy Boistel, 'Training Seafarers in Astronomy: Methods, Naval Schools and Naval Observatories in 18th- and 19th-century France', in, *The Heavens on Earth: Observatories and Astronomy in Nineteenth-Century Science and Culture*, ed. by David Aubin, Charlotte Bigg and H. Otto Sibum (Durham, NC: Duke University Press, 2010), pp. 148–73.

[32] Archives nationales, fonds Marine, 2 JJ 69 (Delisle's papers), fol. 109.

[33] Boistel, 'L'astronomie nautique', part II, especially pp. 160–2.

[34] Jefferson dos Santos Alves, 'O Planetario Lusitano de Eusébio da Veiga e a Astronomia em Portugal no século XVIII', *Revista Brasileira de História da Ciência*, 6 (2013), 340.

[35] Jim Bennett, 'The Travels and Trials of Mr Harrison's Timekeeper', in *Instruments, Travel and Science. Itineraries of Precision from the Seventeenth to the Twentieth Century*, ed. by Marie-Noëlle Bourguet, Christian Licoppe and Heinz Otto Sibum (London: Routledge, 2002), pp. 75–95.

[36] See, for example, letters between Maskelyne and Jeaurat, National Maritime Museum, Greenwich, REG09/000037 <cudl.lib.cam.ac.uk/view/MS-REG-00009-00037/670> (1775) and <cudl.lib.cam.ac.uk/view/MS-REG-00009-00037/684 and 685> (1789) [accessed 10 February 2015]; see also, Danielle Fauque, 'La correspondance Jérôme Lalande et Nevil Maskelyne: un exemple de collaboration internationale au XVIII$^e$ siècle', in *Jérôme Lalande*, ed. by Boistel, Lamy and Le Lay, pp. 109–28.

[37] Boistel, 'L'astronomie nautique', part II.

[38] Lalande to Flaugergues, Viviers, 14 May 1803 (repeated 27 July), quoted in Simone Dumont and Jean-Claude Pecker, *Jérôme Lalande. Lalandiana I. Lettres à Madame du Pierry et au juge Honoré Flaugergues* (Paris: J. Vrin Dumont, 2007), p. 176: 'Nous nous sommes occupés ces jours-ci à recalculer par les observations les 34 étoiles de Maskelyne dont nous nous servions avec une pleine sécurité, et je trouve 5 à 6 secondes à ajouter aux ascensions droites. En sorte qu'il faudra corriger tous nos catalogues, toutes nos tables et toutes nos longitudes des planètes observées! Ce vieux barbouillon, ivrogne paresseux, avare avait usurpé notre confiance. Il est fort riche, il aurait dû se procurer un calculateur et vérifier plus d'une fois cet important résultat'. Thanks to Simon Schaffer for help with the translation.

[39] 'Copy of a Letter from M. Clairaut to Dr. Bevis, dated Paris, 11 April 1765, from the English Original, in his own Hand', *Gentleman's Magazine*, XXXV (1765), 208.

[40] Boistel, 'L'astronomie nautique', part IV and bibliography.

[41] 'Copy of a letter from M. Clairaut'.

[42] Jean-Le-Rond d'Alembert, 'Recherches sur quelques points d'astronomie physique. De la manière la plus simple de calculer analytiquement & astronomiquement les mouvements de la Lune', *Opuscules Mathématiques*, VI (Paris, David, 1773), mémoire XLV, pp. 1–46 (pp. 43–4): 'On peut meme observer encore que dans les equations que M. Mayer emploie pour ses tables, les valeurs des coefficients ne sont pas exactement les mêmes que celles qu'il a tirées de la théorie ; d'où il paroît résulter que les tables de la M. Mayer ont été dressées en partie sur les observations, par une espèce de tâtonnement, combiné avec les résutats principaux que la théorie fournit.'

[43] Boistel, 'L'astronomie nautique', part IV; see also, Jacques Gapaillard, 'La correspondance astronomique entre l'abbé Nicolas-Louis de Lacaille et Tobias Mayer', *Revue d'histoire des sciences*, 49 (1996), 483–541.

[44] Boistel, 'L'astronomie nautique', part IV for a study of Clairaut's lunar tables.

[45] For another example, see Fernando B. Figueiredo and Guy Boistel, 'José Monteiro da Rocha (1734–1819) and the international debate in the 1760s on astronomical methods to find longitude at sea: its proposals and criticisms of the method of lunar distances of Lacaille' (forthcoming), concerning a manuscript of 1765–66, Biblioteca Nacional de Portugal, Lisbon, Colecção Pombalina, Ms.511. Monteiro da Rocha wrote about lunar-distance methods and his trials on a voyage between Brazil and Portugal. His readings and inspiration were Lacaille, Lalande (longitudes and *Connaissance des temps*) and Maskelyne's *British Mariner's Guide*. See also Fernando Figueiredo, 'José Monteiro da Rocha (1734–1819)', in *The Biographical Encyclopedia of Astronomers*, ed. by Thomas Hockey et al. (2nd edition, Dordrecht: Springer, 2014), pp. 513–15.

[46] Nevil Maskelyne, 'Journal of Voyage to St Helena', 1761, Cambridge University Library, RGO 4/150 <cudl.lib.cam.ac.uk/view/MS-RGO-00004-00150/1> [accessed 10 February 2015].